\documentclass[prb,preprint]{revtex4-1}

\usepackage{amsmath}
\usepackage{amsfonts}
\usepackage{graphicx}
\usepackage{tikz}

\begin{document}

\title{Classical realization of the quantum Deutsch algorithm}

\author{Yohan Vianna$^1$}
\author{Mariana R. Barros$^{1,2}$}
\author{Malena Hor-Meyll$^1$}
\email{malena@if.ufrj.br} 

\affiliation{1- Instituto de F\'isica, Universidade Federal do Rio de Janeiro, Caixa Postal 68528, Rio de Janeiro, Rio de Janeiro, 21941-972, Brazil\\
2 - Departamento de F\'isica, Universidade Federal de Minas Gerais, 31270-901, Belo Horizonte, Minas Gerais, Brazil}

\date{\today}

\begin{abstract}

In the rapidly growing area of quantum information, the Deutsch algorithm is ubiquitous and, in most cases, the first one to be introduced to any student of this relatively new field of research.  The reason for this historical relevance stems from the fact that, although extremely simple, the algorithm conveys all the main features of more complex quantum computations. In spite of its simplicity, the uncountable experimental realizations of the algorithm in a broad variety of physical systems are in general quite involved. The aim of this work is two-fold: to introduce the basic concepts of quantum computation for readers with just a minimum knowledge of quantum mechanics and to present a novel  and entirely accessible implementation of a classical analogue of the quantum Deutsch algorithm. By employing only elementary optical devices, such as lenses and diode lasers, this experimental realization has a striking advantage over all previous implementations: it can easily be understood and reproduced in most basic undergraduate or even high-level school laboratories.  

\end{abstract}

\maketitle

\section{Introduction} 

In the last decades we have been facing an unprecedented revolution in the area of information  and computation theory due to the proposal of quantum information protocols and quantum algorithms. Long standing problems such as the prime factorization of large numbers now have quantum counterparts presenting solution time that scale polynomially with the number of inputs. Among them, the most well-known and simplest algorithm is the Deutsch-Jozsa \cite{Deutsch}.  Given a promise that a certain binary function is either constant or balanced, the aim of this algorithm is to find out in a single run to which set this function belongs. Classically, the simplest algorithm has a solution time that scales exponentially with the number $n$ of input bits, since the function has to be evaluated, in the worst scenario, at least $2^{n-1}+1$ times to assure that it is constant or balanced. Though simple, the Deutsch-Jozsa algorithm encompasses all the main ingredients typically found in most quantum algorithms: all quantum computations are just more complex variations of it \cite{Vedral}. This explains its importance in the field of quantum computation, leading hitherto to an uncountable number of experimental realizations in a variety of physical systems such as nuclear spins \cite{Lloyd}, ion traps \cite{Gulde}, quantum dots \cite{Bianucci}, superconducting devices \cite{DiCarlo}, electronic spins \cite{Shi}, photonic degrees of freedom \cite{Zhang} and macroscopic ensembles \cite{Byrnes}, to name a few. Despite the fact that the algorithm is composed by a small number of elementary logical gates, such implementations can be highly involved, depending on which physical system is being used for the experimental realization. The aim of this paper is to present a very simple implementation of the Deutsch algorithm, a special case of the Deutsch-Jozsa algorithm when $n=1$, based on quite simple optical devices such as lenses and diode  lasers, suitable for a first understanding of the underlying principles of quantum computing for undergraduate students. Our optical implementation relies on a very recent proposal \cite{Larsson} which exploits a toy model \cite{Spekkens} for constructing a classical analogue for the Deutsch-Jozsa algorithm that, nevertheless, captures all its relevant quantum features.  

The paper is organized as follows. In Section II  we briefly discuss some useful concepts for describing a quantum algorithm such as quantum bits and quantum gates. After this brief introduction, we present in Section III the original quantum Deutsch algorithm.  We then proceed in Section IV with the description of the classical analogue based on the aforementioned toy model. In Section V we detail the implementation of the analogue version of the Deutsch algorithm using an all-optical setup. We conclude and present future perspectives in Section VI. 

\section{A brief discussion on quantum bits and quantum gates}

Any quantum algorithm can be described by a graphical diagram, called {\it circuit}, composed by a set of quantum logical gates performing unitary operations upon a certain number of input quantum bits (qubits).  Usually, the solution of the algorithm is given by one or more outputs, which are result of the overall gate operations on the input qubits. In order to be unitary, a quantum circuit, opposite to a classical one, must have the same number of inputs and outputs.  Fig.~\ref{FIG1} illustrates a typical quantum circuit, which should be evaluated from left to right:
\begin{figure}[ht]
	\centering
	\includegraphics[width= 8cm]{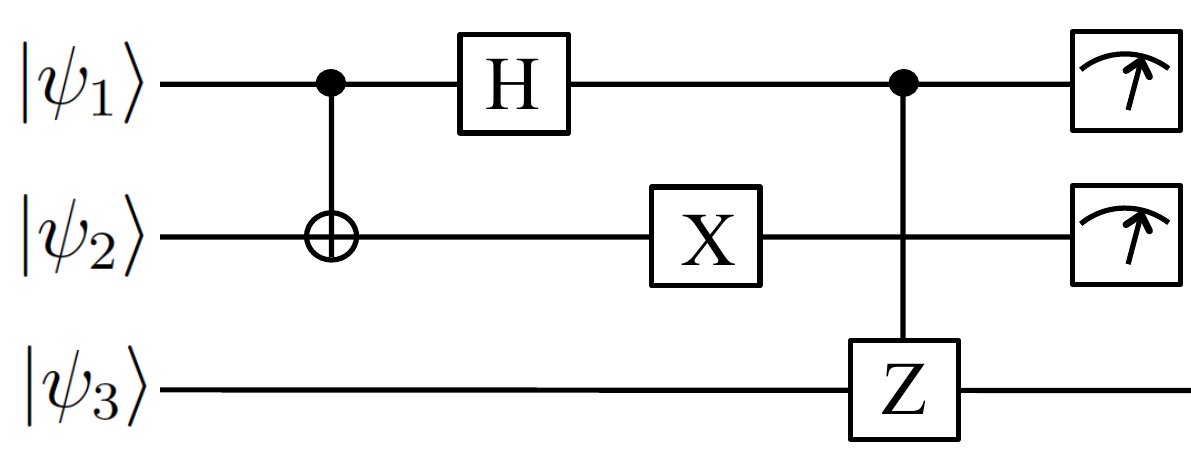}
	\caption{A typical quantum circuit.}
	\label{FIG1}
\end{figure}  

\begin{figure}[ht]
	\centering
	\includegraphics[width= 8cm]{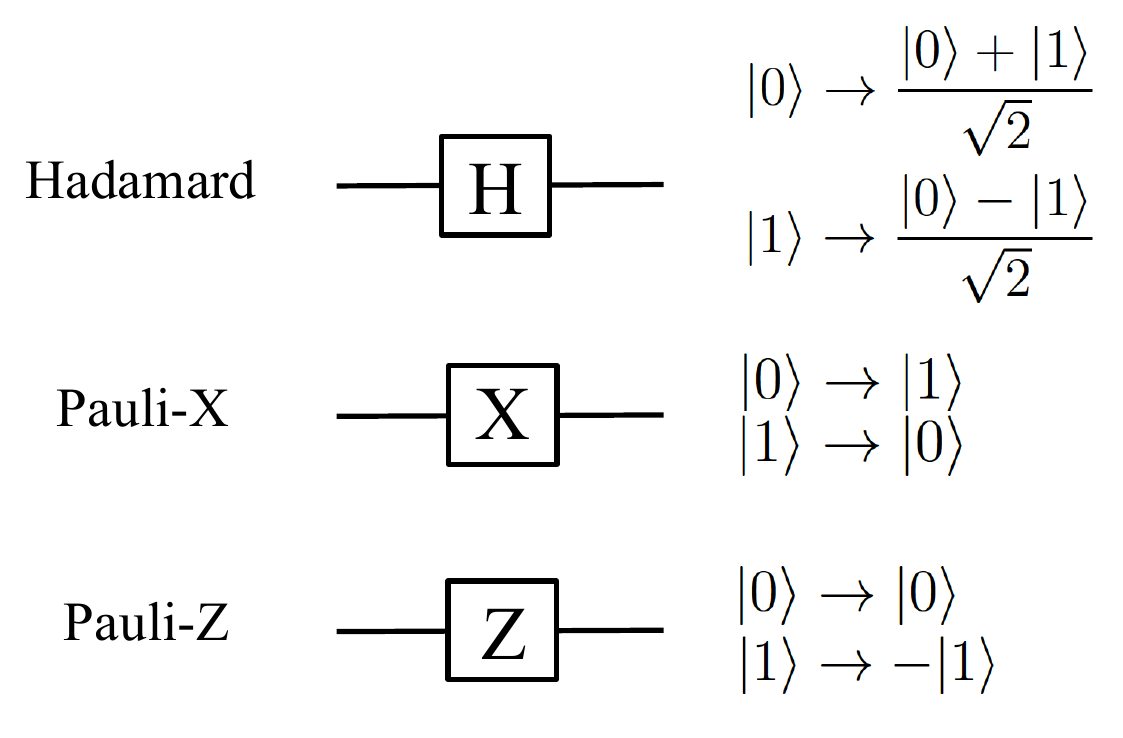}
	\caption{Examples of one-qubit logical quantum gates and their transformation upon the states of the computational basis.}
	\label{FIG2}
\end{figure}

 Horizontal lines represent qubits and symbols or boxes upon these lines correspond to logical gates. Projective measurements, performed in output qubits, are  commonly represented by the ``meter'' symbols as the ones shown at the end of the first two lines in Fig.~\ref{FIG1}. A qubit can be implemented by any binary quantum physical system. We shall label these states as $|0\rangle$ and $|1\rangle$. By convention, these states form an orthonormal basis which is called the {\it computational basis}.  In this basis, the most general state of a qubit can be written as:
\begin{eqnarray}
|\psi\rangle = \cos \left(\dfrac{\theta}{2}\right) |0\rangle + e^{i\phi}\sin\left(\dfrac{\theta}{2}\right) |1\rangle,
\end{eqnarray}
where $\theta \in [0,\pi]$ and $\phi \in [0,2\pi)$.
The most elementary logical quantum gates are those which act upon a single qubit. To illustrate such gates, we choose those used in the Deutsch algorithm, for future convenience. Due to the linearity of quantum mechanics, it suffices to show how these gates act on basis states.  The symbols  for the quantum logical gates called {\it Hadamard} (H), {\it Pauli-X} (X) and {\it Pauli-Z} (Z) and their unitary transformations upon the computational states are shown in Fig.~\ref{FIG2}.

Considering that in quantum mechanics the multiplication by a global phase does not change a physical state, it is clear that the Pauli-Z gate does not change states belonging to the computational basis. Yet, it can significantly change the state of a coherent superposition of basis states. For instance, taking a state to an orthogonal one as in: 

\begin{eqnarray}
Z\; \frac{|0\rangle + |1\rangle}{\sqrt{2}} = \frac{|0\rangle - |1\rangle}{\sqrt{2}};  \nonumber\\
Z \;\frac{|0\rangle - |1\rangle}{\sqrt{2}} = \frac{|0\rangle + |1\rangle}{\sqrt{2}}.
\end{eqnarray}

\begin{figure}[ht]
	\centering
	\includegraphics[width= 8cm]{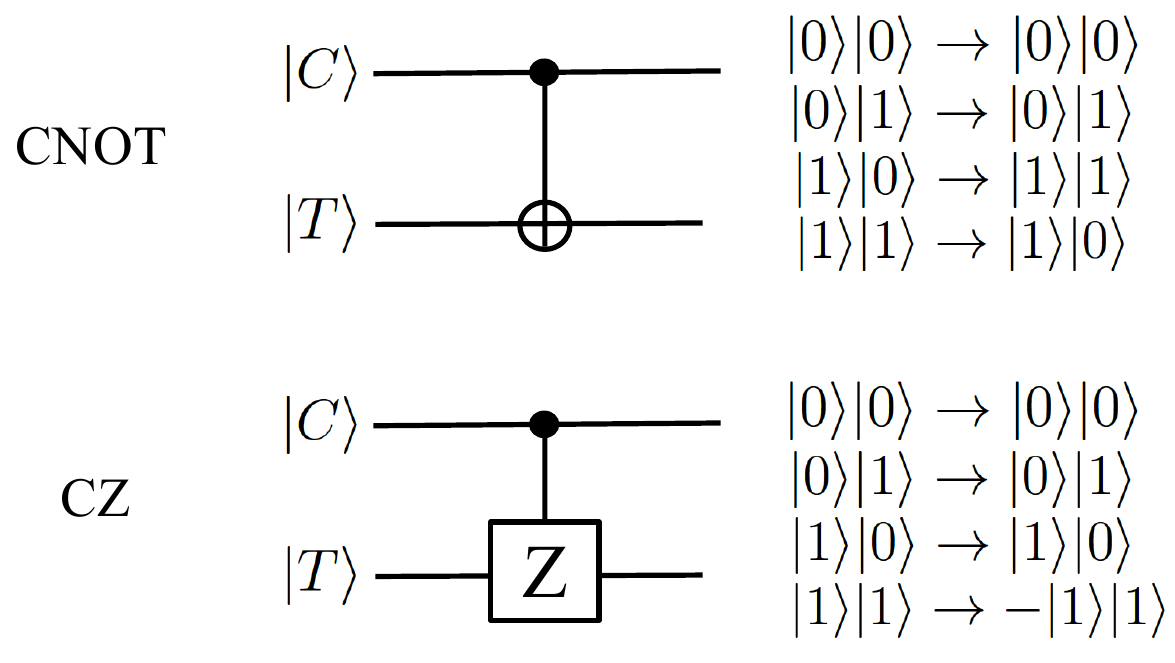}
	\caption{Examples of two-qubit logical quantum gates.}
	\label{FIG3}
\end{figure}

To illustrate two-qubit logical gates we introduce the {\it controlled-NOT} (CNOT) and {\it controlled-Z} (CZ) gates, which are also employed in the Deutsch algorithm. Fig.~\ref{FIG3} shows their respective symbols followed by the description of their unitary transformations on the computational basis for two qubits $\{|0\rangle|0\rangle, |0\rangle|1\rangle, |1\rangle|0\rangle,|1\rangle|1\rangle\}$.
The state of the first qubit [represented by the line with the symbol $\bullet$,  known as the {\it control} qubit (C)] determines the final state of the other qubit  [referred as  {\it target} qubit (T)]: whenever $|C\rangle=|0\rangle$, the target qubit remains unchanged; conversely, when $|C\rangle= |1\rangle$, a single qubit operation is applied on the target qubit. For the CNOT, the single qubit operation corresponds to a Pauli-X gate. Note that, in the computational basis, $X|x\rangle = |1\oplus x\rangle$, where $\oplus$ means addition modulo 2. This means that in this basis the CNOT can be written as:
\begin{eqnarray}
CNOT(|C\rangle|T\rangle) =  |C\rangle|C\oplus T\rangle.
\end{eqnarray}
This explains the symbol $\oplus$ on the target qubit, traditionally used in the representation of the CNOT gate, instead of the usual symbol for the Pauli-X gate.
In the same way, for the CZ gate, the Pauli-Z gate should be applied to the target qubit whenever the control is set to $|1\rangle$. In the computational basis the control qubit is unaffected by both transformations.

\begin{figure}[ht]
	\centering
	\includegraphics[width= 8cm]{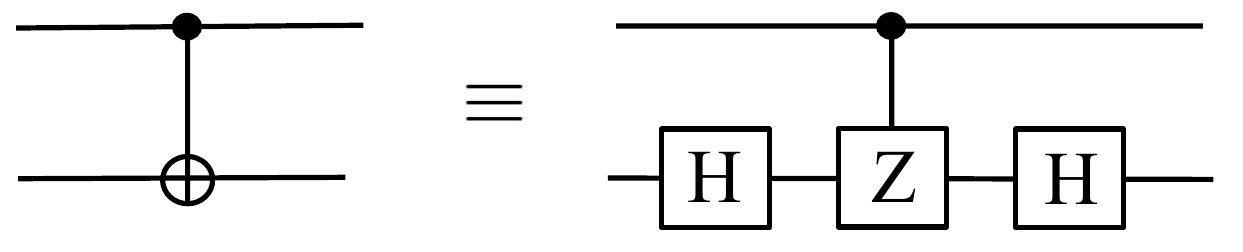}
	\caption{CNOT gate implemented with a CZ and two Hadamard gates.}
	\label{FIG4}
\end{figure}

It is easy to demonstrate that the CNOT gate can be decomposed into the following sequences of more elementary gates shown in Fig.~\ref{FIG4}.

This decomposition will prove to be useful for our implementation.

Gates of an arbitrary number of qubits are defined in similar ways. The concepts outlined above allows us to proceed to the description of the original Deutsch-Jozsa algorithm. Detailed discussion on more general gates and algorithms can be found in the literature \cite{Chuang}.

\section{The original quantum Deutsch algorithm}
\begin{figure}[ht]
	\centering
	\includegraphics[width= 10cm]{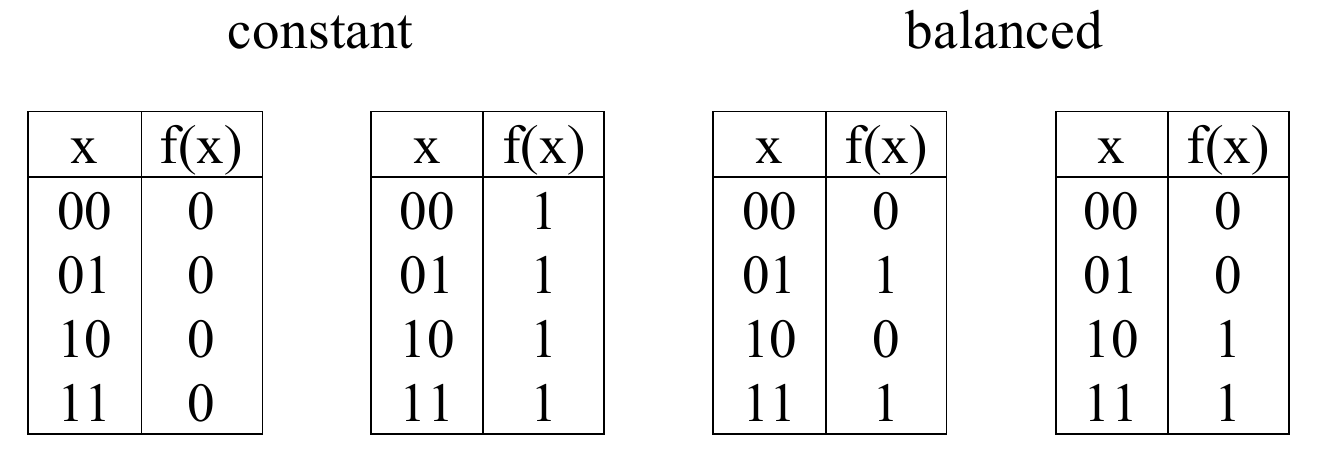}
	\caption{Examples of constant and balanced functions for $n=2$.}
	\label{FIG5}
\end{figure}

The goal of the algorithm proposed by David Deutsch and Richard Jozsa in 1992 is to determine whether an unknown binary function $f(x)$, defined by $f : \left\{0,1\right\}^{n}  \rightarrow \left\{0,1\right\}$, where $n \in \mathbb{N}^*$ is the number of input bits, is constant (equal to $1$ or $0$ for all values of $x$) or balanced (equal to $1$ for exactly half of the values of $x$ and $0$ for the other half). The algorithm relies on the premise that only functions of these two types will be analyzed. Fig.~\ref{FIG5} shows some examples of constant and balanced functions for $n=2$.

In a straightforward classical algorithm, one evaluates $f(x)$ for a single value of $x$ at a time. In the worst situation, to conclude whether the function is constant or balanced one needs to perform the operation $2^{n-1} +1$ times, namely, half the number of valid combinations of input bits plus one. 

However, if this problem is solved using quantum resources, as prescribed by the Deutsch-Jozsa algorithm, it is possible to determine whether the function is constant or balanced evaluating it globally in a single run, exploiting the possibility of a coherent quantum superposition of states as input. Therefore, the quantum algorithm allows solving the problem in exponentially less time if compared to the classical procedure.

\begin{figure}[ht]
	\centering
	\includegraphics[width= 5.5cm]{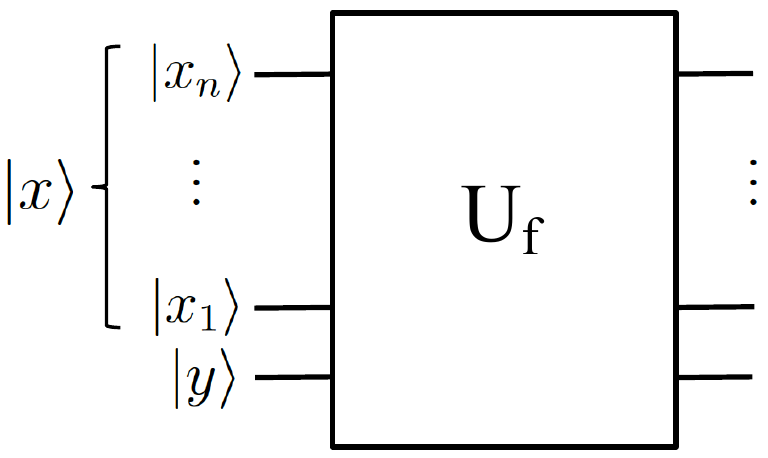}
	\caption{The oracle of the Deutsch-Jozsa algorithm.}
	\label{FIG6}
\end{figure}

Being $f(x)$ a constant or balanced function, it is impossible to devise a unitary operator (and hence a reversible one) to evaluate $f(x)$ by using just $|x\rangle$ as input. This problem can be circumvented by the introduction of an auxiliary qubit. So,
for the sake of unitarity, the quantum algorithm relies on an operator $U_f$, known as {\it oracle}, through which the function can be evaluated.  The oracle takes an input $|x\rangle$, called {\it register}, composed of $n$ qubits and an auxiliary input $|y\rangle$, called {\it ancilla}, composed by just a single qubit, as shown in Fig.~\ref{FIG6}.  

The corresponding outputs are related to the inputs according to the unitary operation $U_{f}$ given by:
\begin{eqnarray}
U_f (|x\rangle|y\rangle) = |x\rangle|f(x)\oplus y\rangle,
\end{eqnarray}
where $|x\rangle \equiv |x_{n}\rangle... |x_1\rangle$. If $|y\rangle=|0\rangle$, the final state of the ancilla is $|f(x)\rangle$, thus for each function $f$ we have a different oracle. The advantage of the quantum algorithm is that interrogating the oracle only once is sufficient to determine the type of the corresponding function. 

For illustration and future convenience we will show the oracle circuits corresponding to the four functions of the Deutsch algorithm ($n=1$), namely, the constant functions $f(x)=0$ and $f(x)=1$ and the balanced ones $f(x)=x$ and $f(x)=\bar{x}$, where $\bar{x} = 1 \oplus x$, flipping the value of $x$ (Pauli-X operation). In this particular situation the register, composed by a single qubit, will be referred simply as $|x\rangle$. 

\begin{figure}[ht]
	\centering
	\includegraphics[width= 8cm]{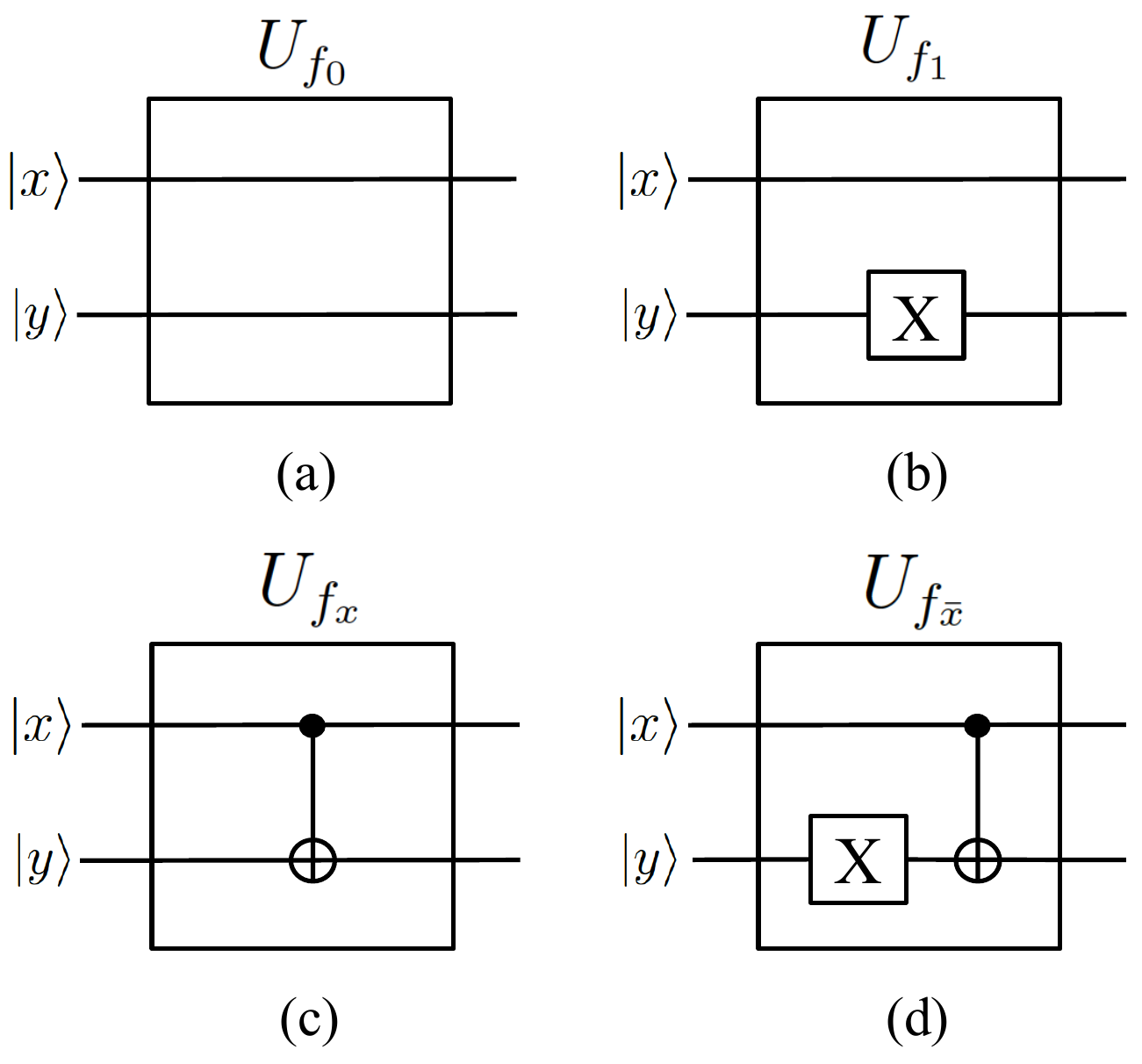}
	\caption{Oracle circuits for the Deutsch algorithm.}
	\label{FIG7}
\end{figure}

The simplest oracle is the one corresponding to the constant function $f(x)=0$, which will be labelled $f_0$ hereafter:
\begin{eqnarray}
U_{f_0} (|x\rangle|y\rangle) = |x\rangle|f(x)\oplus y\rangle =  |x\rangle|0\oplus y\rangle = |x\rangle|y\rangle. 
\end{eqnarray}
So, in this case, the oracle turns out to be the identity operator, as shown in Fig.~\ref{FIG7}(a).

The constant function $f(x)=1$, which will be labelled $f_1$, gives rise to the following oracle transformation:
\begin{eqnarray}
U_{f_1} (|x\rangle|y\rangle) =  |x\rangle|1\oplus y\rangle = |x\rangle X|y\rangle, 
\end{eqnarray}
whose circuit is shown in Fig.~\ref{FIG7}(b). 

The balanced function $f(x) = x$, which will be referred as $f_x$,  is implemented by the oracle corresponding to the following transformation:
\begin{eqnarray}
U_{f_x} (|x\rangle|y\rangle) =  |x\rangle|x\oplus y\rangle = CNOT (|x\rangle|y\rangle), 
\end{eqnarray}
being $|x\rangle$ the control and $|y\rangle$ the target in the CNOT gate. So, in this case, the oracle function coincides with the CNOT gate as shown in Fig.~\ref{FIG7}(c). Finally, the balanced function $f(x) = \bar{x}$, which we will identify by $f_{\bar{x}}$, leads to the following oracle operator:
\begin{eqnarray}
U_{f_{\bar{x}}} (|x\rangle|y\rangle) =  |x\rangle|\bar{x}\oplus y\rangle = |x\rangle|x\oplus \bar{y}\rangle = CNOT (|x\rangle X |y\rangle). 
\end{eqnarray}
Fig.~\ref{FIG7}(d) shows the circuit for $f_{\bar{x}}$ oracle.

\begin{figure}[ht]
	\centering
	\includegraphics[width= 8cm]{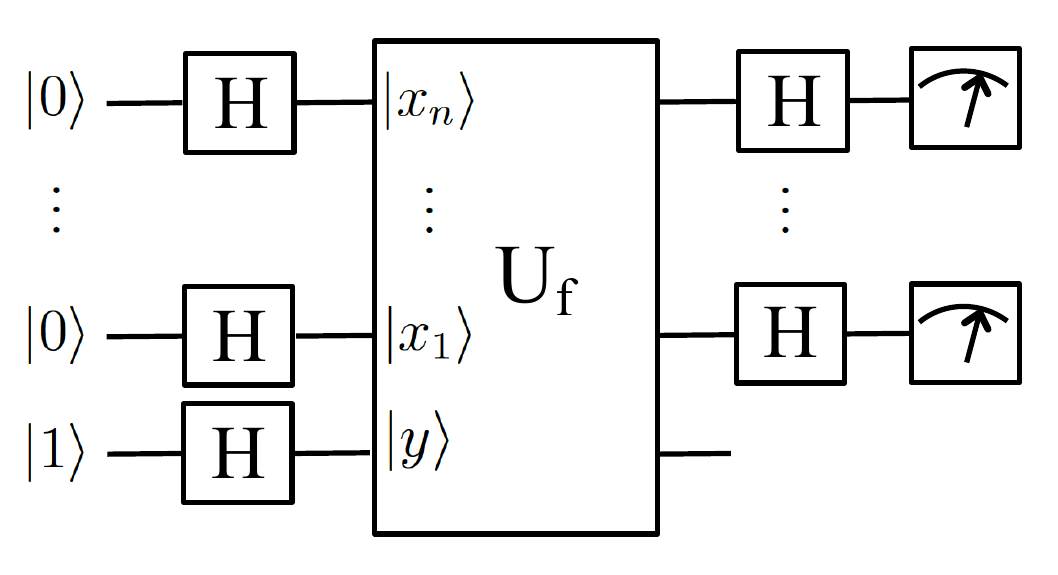}
	\caption{Quantum circuit of the Deutsch-Jozsa algorithm.}
	\label{FIG8}
\end{figure}

Oracles for functions with a greater number of inputs are obtained in a similar manner. 
The complete circuit representing the Deutsch-Jozsa algorithm for any function with an arbitrary number of inputs $n$ is shown in Fig.~\ref{FIG8}.

\begin{figure}[ht]
	\centering
	\includegraphics[width= 8cm]{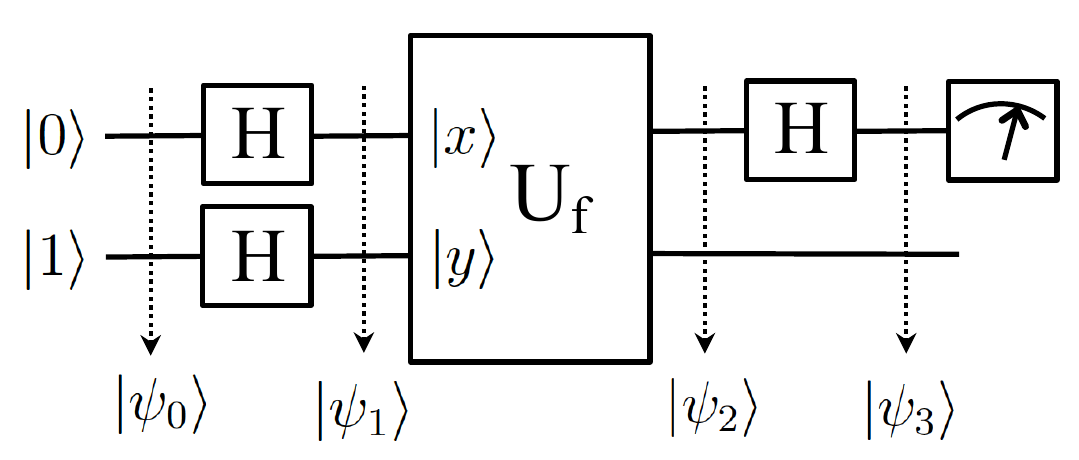}
	\caption{Quantum circuit of the Deutsch algorithm.}
	\label{FIG9}
\end{figure}

For the sake of simplicity, we will analyze the problem for a one-qubit register ($n=1$), shown in Fig.~\ref{FIG9}, which is simply called Deutsch algorithm, as mentioned before. 

The generalization for $n$-qubits, though more involved, is achieved by a similar procedure \cite{Chuang}. Since for $n=1$ there are two types of constant and two types of balanced functions, the goal is to determine which type an unknown function $f(x)$ belongs to. To  explain how the algorithm works, we shall proceed into a step-by-step evaluation of the states throughout the circuit. The input state is chosen to be:
\begin{eqnarray}
|\psi_0\rangle = |0\rangle|1\rangle,
\end{eqnarray}
where the first qubit stands for the register and the second one for the ancilla. After the Hadamard gates, this state evolves to
\begin{eqnarray}
|\psi_1\rangle =H|0\rangle H|1\rangle=\frac{(|0\rangle+|1\rangle)}{\sqrt{2}}\frac{(|0\rangle-|1\rangle)}{\sqrt{2}} =\frac{1}{\sqrt{2}}\left[ |0\rangle \frac{(|0\rangle-|1\rangle)}{\sqrt{2}}+ |1\rangle \frac{(|0\rangle-|1\rangle)}{\sqrt{2}}\right].
\label{A}
\end{eqnarray}
The RHS of Eq.~\eqref{A} will be proved to be convenient for later evaluation of the oracle action on $|\psi_1\rangle$. Indeed, to simplify understanding of the oracle transformation, let us first consider the action of $U_f$ in the following state: 
\begin{eqnarray}
U_f |0\rangle \frac{|0\rangle - |1\rangle}{\sqrt{2}} &=& \frac{1}{\sqrt{2}} [|0\rangle|0 \oplus f(0)\rangle - |0\rangle |1 \oplus f(0)\rangle].
\end{eqnarray}
Realizing that we have two possible values for $f(0)$, we get:
\begin{eqnarray}
U_f |0\rangle \frac{|0\rangle - |1\rangle}{\sqrt{2}} =
\begin{cases}
\frac{1}{\sqrt{2}} [|0\rangle|0\rangle - |0\rangle |1\rangle]= |0\rangle \frac{|0\rangle - |1\rangle}{\sqrt{2}} \;\;\;\;\;\;\;\mbox{for}\;\;f(0)=0,\\
\frac{1}{\sqrt{2}} [|0\rangle|1\rangle - |0\rangle |0\rangle] = -|0\rangle \frac{|0\rangle - |1\rangle}{\sqrt{2}} \;\;\;\;\mbox{for}\;\;f(0)=1.
\label{B}
\end{cases}
\end{eqnarray}
Hence we can rewrite Eq.~\eqref{B} as:
\begin{eqnarray}
U_f |0\rangle \frac{|0\rangle - |1\rangle}{\sqrt{2}} = (-1)^{f(0)} |0\rangle \frac{|0\rangle - |1\rangle}{\sqrt{2}}.
\label{C}
\end{eqnarray}
In the very same way one can easily conclude that:
\begin{eqnarray}
U_f |1\rangle \frac{|0\rangle - |1\rangle}{\sqrt{2}} = (-1)^{f(1)} |1\rangle \frac{|0\rangle - |1\rangle}{\sqrt{2}}.
\label{D}
\end{eqnarray}
We are now ready to obtain state $|\psi_2\rangle$ resulting from the action of the oracle in state $|\psi_1\rangle$:
\begin{eqnarray}
|\psi_2\rangle = U_f |\psi_1\rangle &=& U_f \frac{1}{\sqrt{2}} \left[|0\rangle \frac{(|0\rangle-|1\rangle)}{\sqrt{2}}+ |1\rangle \frac{(|0\rangle-|1\rangle)}{\sqrt{2}}\right]\nonumber\\
&=&  \left[\frac{(-1)^{f(0)} |0\rangle + (-1)^{f(1)}|1\rangle}{\sqrt{2}}\right] \frac{(|0\rangle-|1\rangle)}{\sqrt{2}},
\end{eqnarray}
using Eqs~\eqref{C} and \eqref{D}. Depending on whether $f(0)=f(1)$ (constant function) or $f(0)\ne f(1)$ (balanced function) we are left with:
\begin{eqnarray}
|\psi_2\rangle = \begin{cases}
(-1)^{f(0)} \frac{|0\rangle+|1\rangle}{\sqrt{2}}\frac{|0\rangle-|1\rangle}{\sqrt{2}}\;\;\;\;\mbox{for}\;\; f(0)=f(1),\\
(-1)^{f(0)}  \frac{|0\rangle-|1\rangle}{\sqrt{2}}\frac{|0\rangle-|1\rangle}{\sqrt{2}}\;\;\;\;\mbox{for}\;\; f(0)\ne f(1).
\end{cases}
\end{eqnarray}
The final state $|\psi_3\rangle$ is obtained applying a Hadamard gate on the first qubit:
\begin{eqnarray}
|\psi_3\rangle = \begin{cases}
(-1)^{f(0)} |0\rangle \frac{|0\rangle-|1\rangle}{\sqrt{2}}\;\;\;\;\mbox{for}\;\; f(0)=f(1),\\
(-1)^{f(0)} |1\rangle \frac{|0\rangle-|1\rangle}{\sqrt{2}}\;\;\;\;\mbox{for}\;\; f(0)\ne f(1).
\end{cases}
\end{eqnarray}
Finally, the state of the first qubit indicates the type of function $f(x)$: $|0\rangle$ for constant functions and $|1\rangle$ for the balanced ones. Note that the unitary operation corresponding to the oracle was evaluated only once, which contrasts with the classical procedure that would necessarily demand two evaluations. This advantage comes from the fact that a quantum computer can evaluate both $f(0)$ and $f(1)$ simultaneously. Clearly the gain is exponentially greater as the number $n$ of register qubits increases. 

The simplicity of the Deutsch algorithm rendered it the privilege of being the first quantum algorithm to be experimentally demonstrated \cite{Jones}.

In the next section we show how a classical analogue of the Deutsch circuit can be obtained as a direct map of the quantum version.

\section{The classical Deutsch algorithm}
\label{CDA}

Based on the toy model proposed by Spekkens \cite{Spekkens}, a classical analogue of the Deutsch-Jozsa algorithm was formulated  by Johansson and Larsson \cite{Larsson}.  In their proposal, the qubit states and the quantum logical gates of the original algorithm are simply substituted by the equivalent ones defined by the toy model. 
In the subsection \ref{subsection:toymodel} we present a short review of Spekkens' toy model and in subsection  \ref{subsection:analogue} we describe the classical analogue of the Deutsch algorithm. 

\subsection{The toy model}
\label{subsection:toymodel}

As a modeling theory, the toy theory was thought as a classical model to describe two different views for a physical system state. When studying the most basic principles of classical mechanics, one faces the description of objects as particles, with a well-established velocity and position in a specific time (or inside an interval of time). That concept, of a particle in the space with position and velocity vectors associated to them as functions of time, gives us a complete knowledge of the system for any time t. This idea of state that measures the reality of a given system will be called here an \textit{ontic} state. On the other hand, if one focuses his attention into the probabilistic nature of statistical mechanics, the abstract notion of probability distribution over a given space shall be confronted. This notion represents the limits of what the agent performing an experiment knows about the system: it is not a state of reality of the system, but a state of knowledge, by the agent, about the system. We will call this an \textit{epistemic} state. In the toy model, a restricted class of quantum states is represented by means of an epistemic state, devised as a combination of ontic states. 

\begin{figure}[ht]
	\centering
	\includegraphics[width= 13cm]{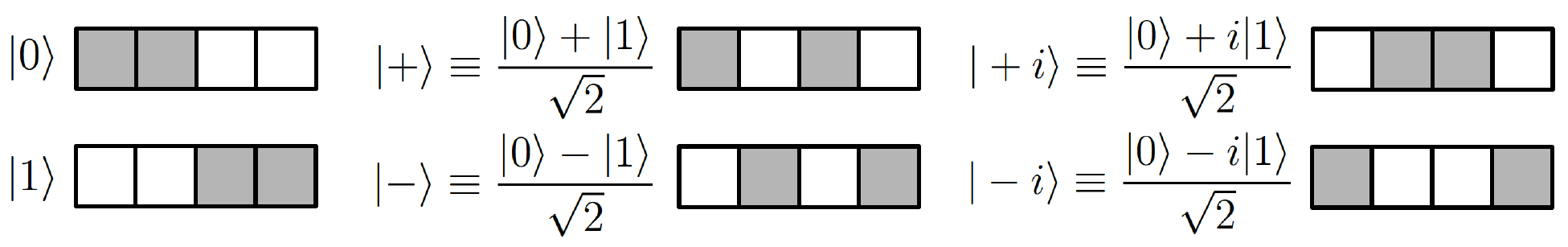}
	\caption{Representation of epistemic states of one-qubit through four ontic states.}
	\label{FIG10}
\end{figure}

Let us begin with the description of the analogues of one-qubit states. The toy model is able to represent classical analogues of all eigenstates of Pauli matrices. Four classical bits, understood in this context as ontic states, will be used in convenient configurations to represent single-qubit epistemic states, as depicted by the graphical diagrams shown in Fig.~\ref{FIG10}.
	
\begin{figure}[ht]
	\centering
	\includegraphics[width= 13cm]{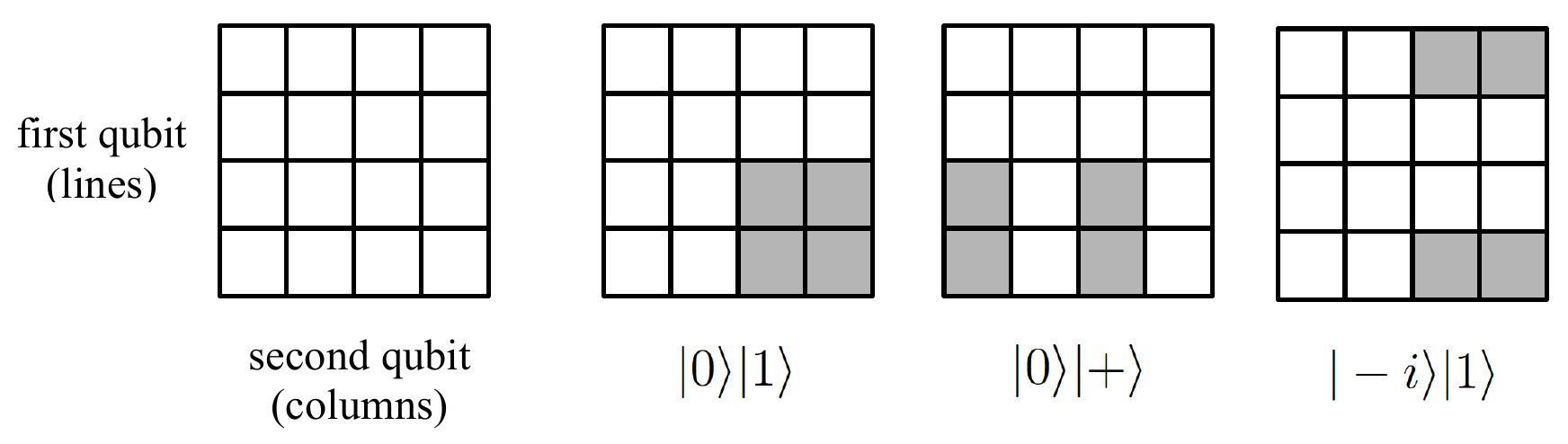}
	\caption{Examples of epistemic states of two qubits.}
	\label{FIG11}
\end{figure}

It is noteworthy that other geometric configurations could be chosen to represent these states. The convention used here is to represent one qubit as a virtual line grid composed by classical bits which could be {\it on} (grey square) or {\it off} (white square). According to the toy model, we can extend this diagram to represent two qubits by means of a four by four virtual matrix grid where the lines represent the ontic states of the first qubit (read from down to up) and the columns (read from left to right) represent the ontic states of the second qubit, as illustrated in Fig.~\ref{FIG11}.

\begin{figure}[ht]
	\centering
	\includegraphics[width= 8cm]{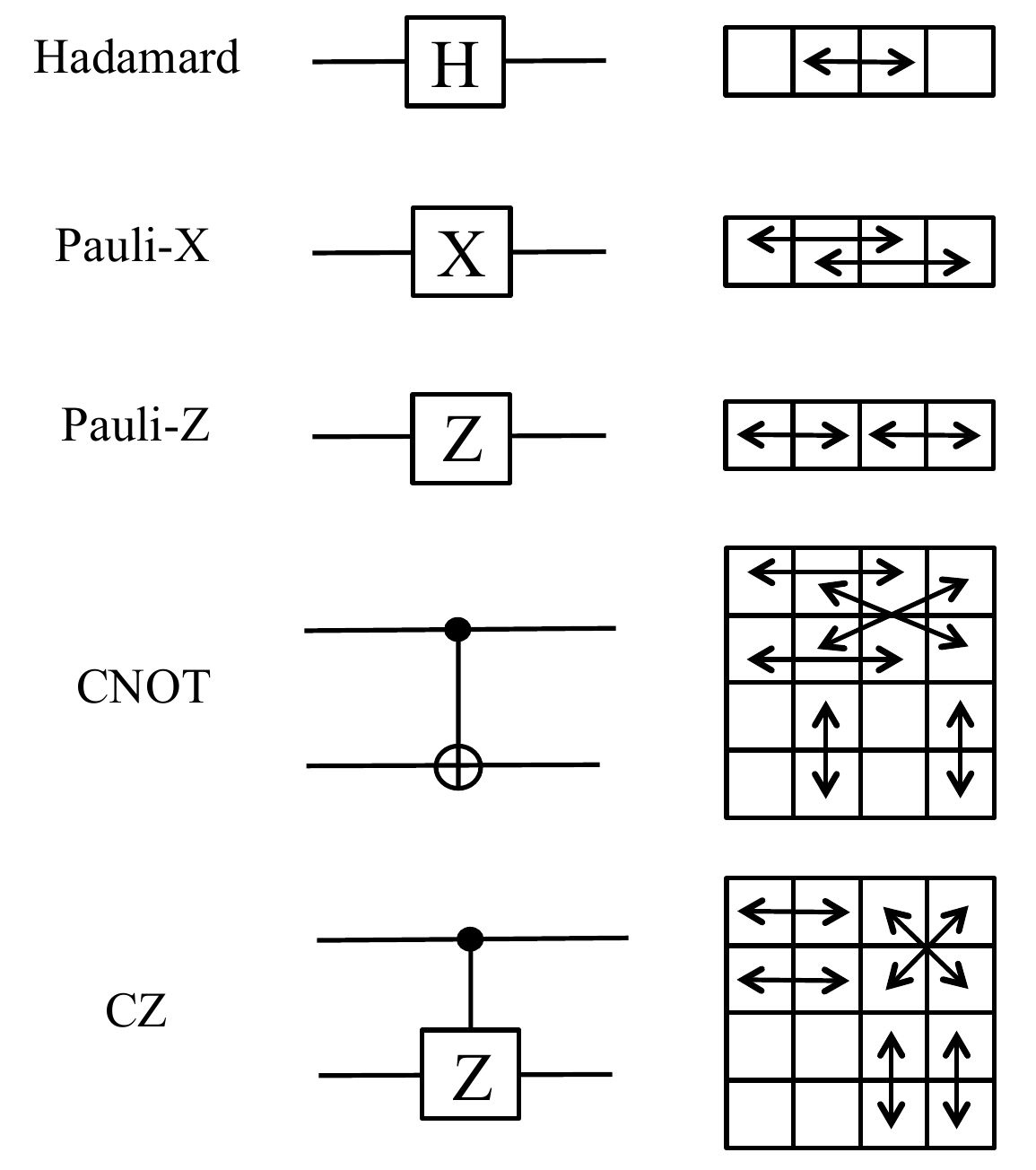}
	\caption{Permutations corresponding to Hadamard, Pauli-X, Pauli-Z, CNOT and CZ gates.}
	\label{FIG12}
\end{figure}

Given these representations for qubits, the quantum gates analogues arise naturally as permutations of the classical bits. The permutations corresponding to the one-qubit Hadamard, Pauli-X and Pauli-Z gates, as well as those corresponding to the two-qubit controlled-NOT (CNOT) and controlled-Z (CZ) gates are shown in Fig.~\ref{FIG12}.

\begin{figure}[ht]
	\centering
	\includegraphics[width= 13cm]{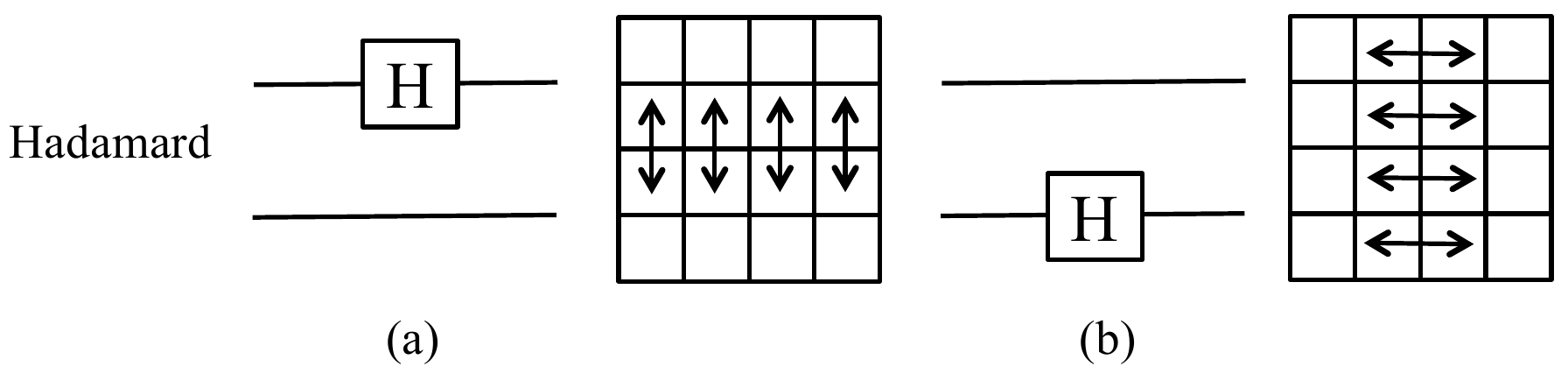}
	\caption{Permutations for the two-qubit configuration of Hadamard gate applied on (a)~the  first  and (b)~second qubit.}
	\label{FIG13}
\end{figure}

In the implementation of the Deutsch algorithm, we will be working solely with two-qubit state representation of epistemic states. So it is worth noticing that one-qubit operations extended to two-qubit representation is accomplished by performing the corresponding permutations in lines or columns, depending on whether the operation is applied on the first or the second qubit, respectively. For illustration, we show in Fig.~\ref{FIG13}, the permutations corresponding to the Hadamad gate applied on the first (a) and second (b) qubit. 	

\subsection{The Classical Deutsch Algorithm}
\label{subsection:analogue}

\begin{figure}[ht]
	\centering
	\includegraphics[width= 10cm]{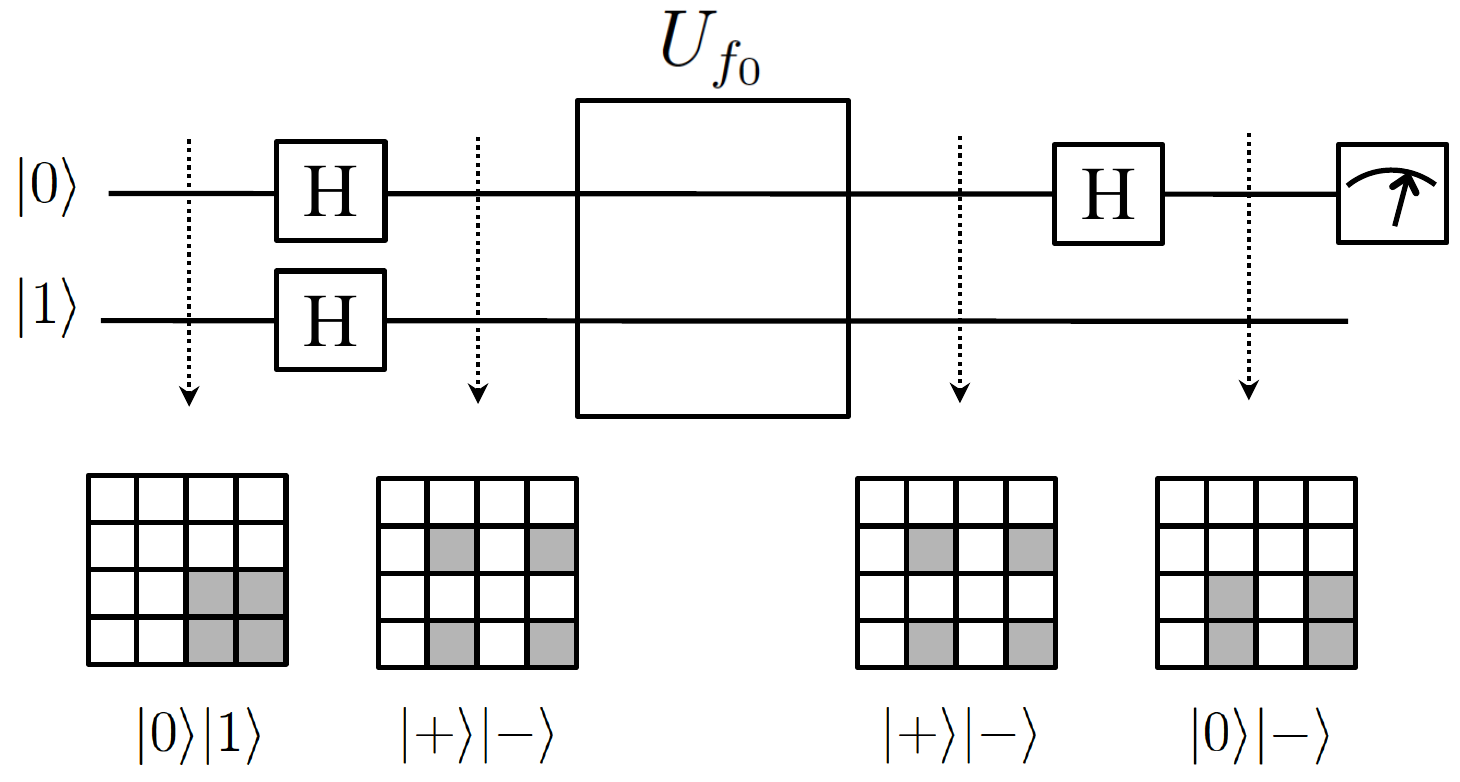}
	\caption{State evolution of the classical Deutsch algorithm for the constant function $f(x)=0$.}
	\label{FIG14}
\end{figure}

We now discuss the classical analogue of the Deutsch algorithm, illustrated in Fig~\ref{FIG9}, within the framework of the toy model.  It consists simply in representing the input state as a four by four grid in the configuration representing $|0\rangle|1\rangle$, the first qubit standing for the register and the second one for the ancilla, and as stated before, replacing each quantum gate by the set of corresponding permutations, as defined in the previous subsection. Hence, execution of the classical analogue of the algorithm for each oracle, shown in Fig.~\ref{FIG7},  will be demonstrated through a step-by-step sequence of the resulting epistemic states after application of each gate. 
For constant function $f(x)=0$,  we end up with the sequence of states shown in Fig.~\ref{FIG14}.

We note that for constant function $f(x)=1$ we will have exactly the same sequence of states shown in Fig.~\ref{FIG14}, since the oracle differs only by a Pauli-X gate which will not affect the state $|+\rangle|-\rangle$, resulting from the action of the Hadamard gates on input state $|0\rangle|1\rangle$. Indeed, we see that in both cases, the final state of the register qubit is $|0\rangle$, as expected for constant functions.

\begin{figure}[ht]
	\centering
	\includegraphics[width= 10cm]{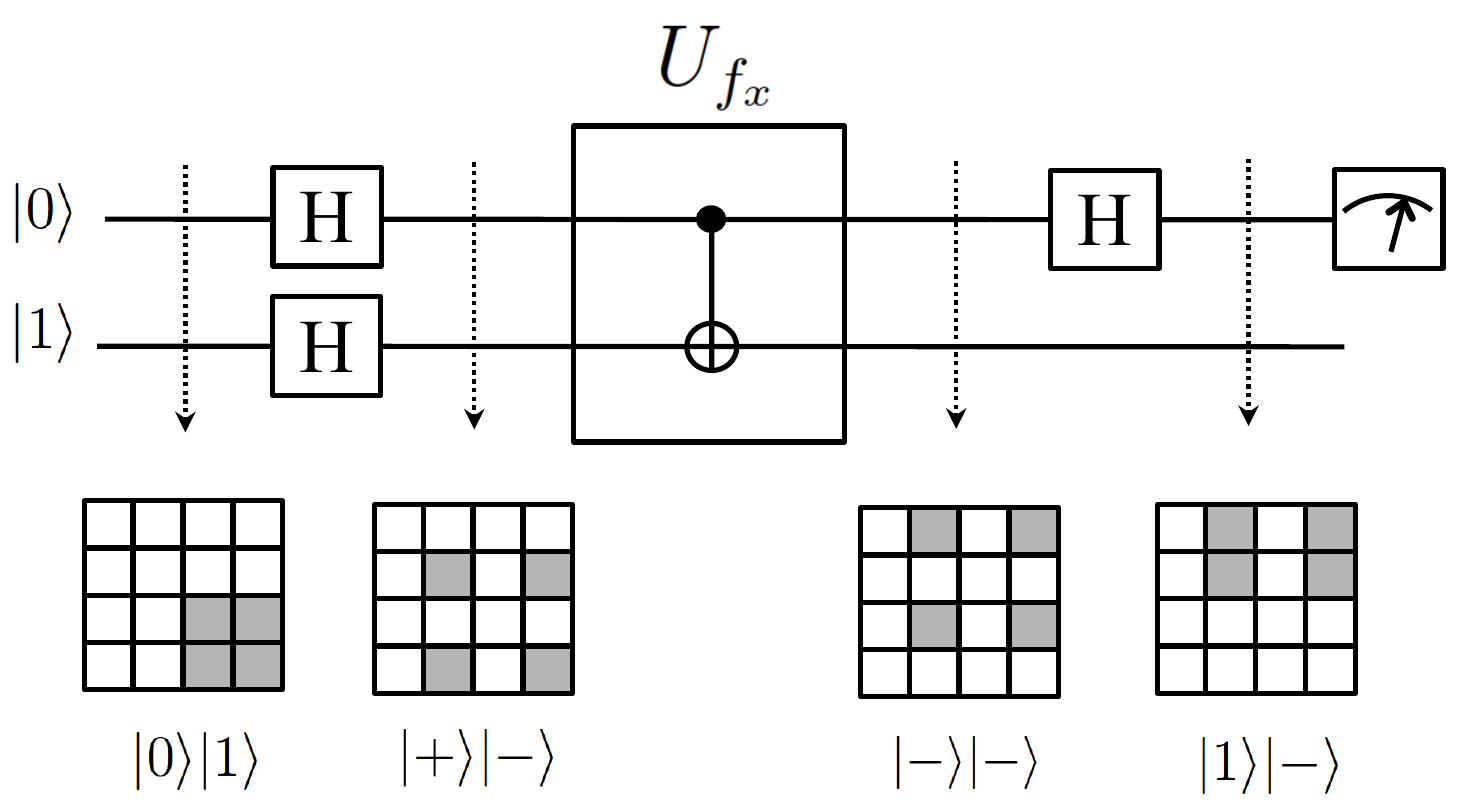}
	\caption{State evolution of the classical Deutsch algorithm for the balanced function $f(x)=x$.}
	\label{FIG15}
\end{figure}

Similarly, for the oracle corresponding to the balanced function $f(x)=x$, we will have the sequence of epistemic states shown in Fig.~\ref{FIG15}.
And, once again, the same sequence of epistemic states shown in Fig.~\ref{FIG15} will be found for the evaluation of the oracle corresponding to the balanced function $f(x)=\bar{x}$, since the introduction of the Pauli-X before the CNOT will have no effect, for the same reason pointed out above. Therefore, for both balanced functions, the final state of the register qubit is $|1\rangle$ as required.

In the next section, we discuss our optical implementation of the Deutsch algorithm following the prescriptions of the toy model, which leads to the behavior just explained.

\section{Optical implementation of the Deutsch algorithm analogue} 
\label{DispEqSection}

The material required for this implementation is usually found in most undergraduate laboratories and even in some high level schools. It consists of diode lasers, spherical and cylindrical convex thin lenses and small pieces of thin glass. 

\subsection{Optical implementation of states and gates}
\label{subsection:implementation}

\begin{figure}[h!]
	\centering
	\includegraphics[scale=0.9]{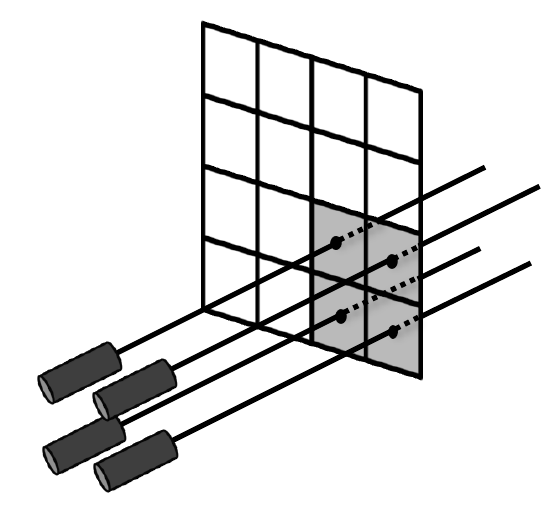}
	\caption{Laser configuration for preparation of the Deutsch  algorithm initial state $|0\rangle|1\rangle$.}
	\label{FIG16}
\end{figure}

The state {\it on} of a classical bit will be represented by a light beam produced by a laser. The absence of a light beam should be understood as a classical bit in the {\it off} state.  For a two-qubit representation, although we will have sixteen possible path beams, each one corresponding to a classical bit, only four will be {\it on} for any state, so we need just four lasers at all. Accordingly, to prepare the initial state $|0\rangle|1\rangle$, the light beams must be disposed in the parallel configuration shown in Fig.~\ref{FIG16}. 
Finally, the grid that serves for the visualization of qubit representation can be drawn on a flat piece of transparent glass.

\begin{figure}[h!]
	\centering
	\includegraphics[scale=0.6]{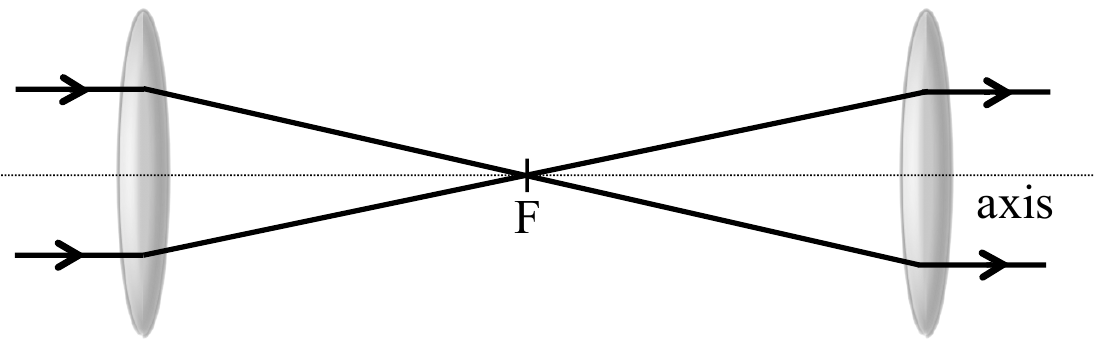}
	\caption{Beam permutation through a pair of confocal spherical convex lenses. F stands for focus.}
	\label{FIG17}
\end{figure}

The permutations required to implement the logical gates are accomplished by combinations of pairs of spherical or cylindrical convex lenses in a confocal configuration. The behavior of a pair of confocal spherical convex lenses in two parallel incoming beams is shown in Fig.~\ref{FIG17}, and illustrates the general idea of the whole experiment to generate the gates permutations.

The overall effect of this lens system is that the outcoming beams remain parallel in comparison with the incoming beams, but emerge spatially interchanged in relation to the optical axis of the lenses.  Although not shown in Fig.~\ref{FIG17}, a beam coincident with the optical axis will remain unaffected by this lens system. This feature will be exploited in our optical implementation of the Pauli-X gate, as we will see.

\begin{figure}[h!]
	\centering
	\includegraphics[scale=0.8]{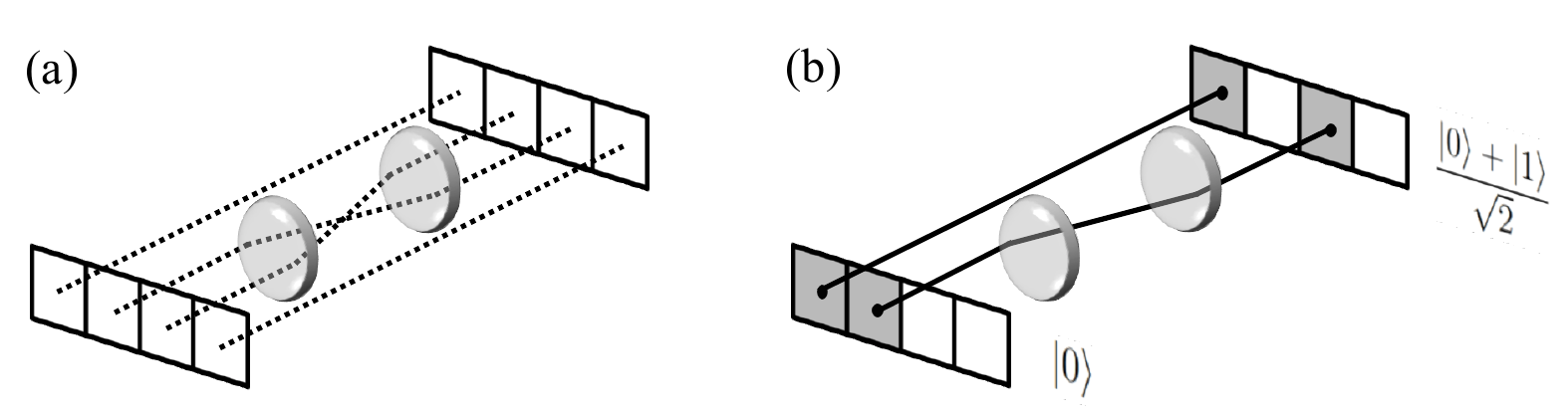}
	\caption{(a) Lens configuration for implementing the Hadamard gate for one qubit. (b) Example showing the action of the lens scheme on the input state $|0\rangle$.}
	\label{FIG18}
\end{figure}

Although not used in the Deutsch algorithm, for pedagogical reasons, let us now discuss the implementation of a single-qubit configuration Hadamard gate using this scheme.  
According to the first part of Fig.~\ref{FIG12}, the Hadamard gate consists in permuting the second and third classical bits. This is accomplished simply by placing a pair of spherical confocal lenses in the beam paths, represented by dotted lines, corresponding to the second and third classical bits, as shown in Fig.~\ref{FIG18}(a). 

For further clarification we show in Fig.~\ref{FIG18}(b) the action of this scheme on the input state $|0\rangle$, where solid lines represents light beams (classical bit in {\it on} state). It is worth exploiting this simple gate to introduce a compact and useful representation to be used in the remainder of this article. In this notation, the pair of confocal lenses of Fig.~\ref{FIG18}(a) will be represented simply by
\begin{figure}[h!]
	\centering
	\includegraphics[scale=0.7]{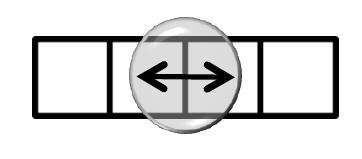}
	\caption{Compact representation of the pair of confocal spherical lens representing a one-qubit Hadamard gate.}
	\label{FIG19}
\end{figure}

\begin{figure}[h!]
	\centering
	\includegraphics[scale=0.7]{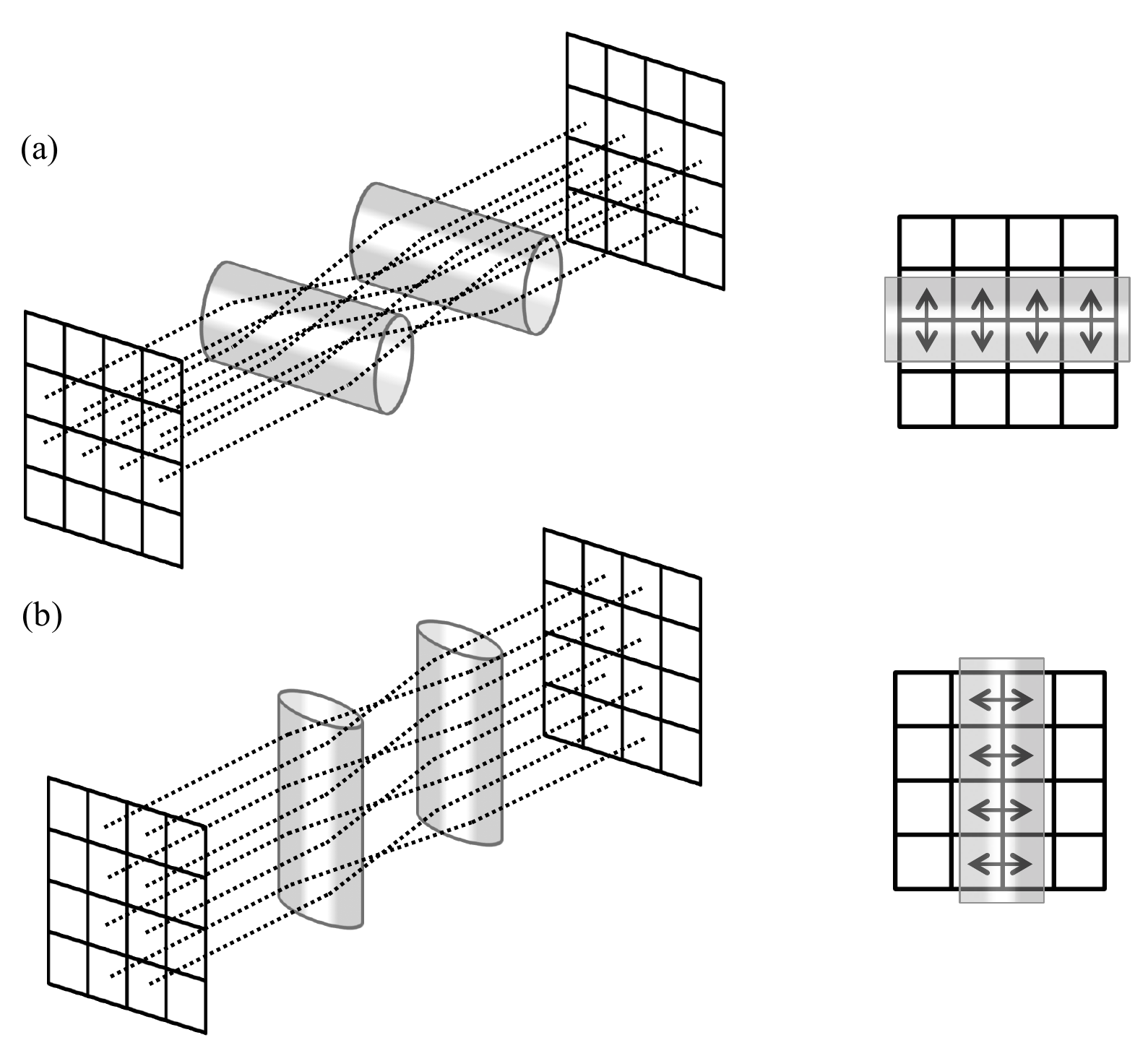}
	\caption{Pair of confocal cylindrical lenses representing a two-qubit configuration Hadamard gate applied on (a)~the first qubit and corresponding compact representation and on (b)~ the second qubit and corresponding compact representation.}
	\label{FIG20}
\end{figure}

Similarly, the implementation of a two-qubit configuration Hadamard gate stands for placing a pair of convex cylindrical confocal lenses in the path of the second and third lines, or columns, depending on whether the Hadamard gate is supposed to be applied on the first or second qubit, respectively. The  longitudinal axis lay parallel to the qubit that it is affected by the gate, such that  the lenses configuration has no effect on the qubit that should remain unchanged. Accordingly, in Fig.~\ref{FIG20}(a)~and~(b), we show the optical implementation of Fig.~\ref{FIG13}(a)~and~(b) and their corresponding compact representations.To avoid an over-polluted picture, we draw only paths corresponding to beams that do suffer changes by the lens system.

\begin{figure}[h!]
	\centering
	\includegraphics[scale=0.7]{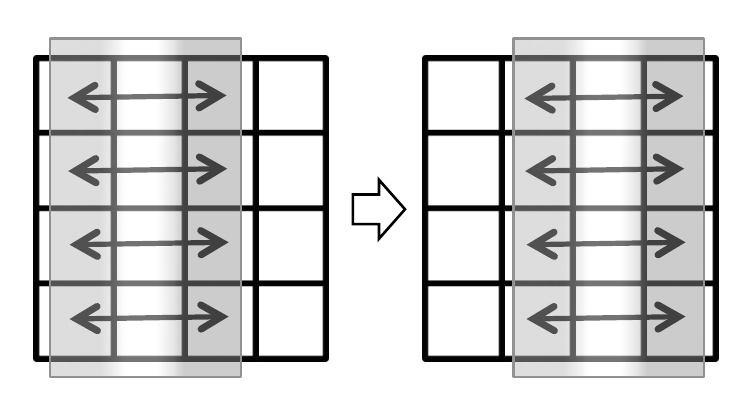}
	\caption{Compact representation of the sequence of two pairs of confocal cylindrical lenses representing a two-qubit configuration Pauli-X gate applied on the second qubit.}
	\label{FIG21}
\end{figure}

From now on we will make use of the compact representation to illustrate the implementation of the remaining gates. For the two-qubit configuration Pauli-X gate applied on the second qubit, as required by the protocol, we use four larger cylindrical lenses in sequence. The first pair permutates the beam paths corresponding to the first and third classical bit columns, since the longitudinal optical axes are set to be coincident with the beam paths corresponding to the second classical bit column, which consequently remains unaffected by the lenses system.  Similarly, the second pair of lenses exchanges the beam paths corresponding to the second and fourth classical bit columns, by making the longitudinal lenses' axes now coincident with the beam path corresponding to the third classical bit column, as shown in Fig.~\ref{FIG21}.
It is worth mentioning that the spatial order of the confocal pairs is irrelevant.

\begin{figure}[h!]
	\centering
	\includegraphics[scale=0.7]{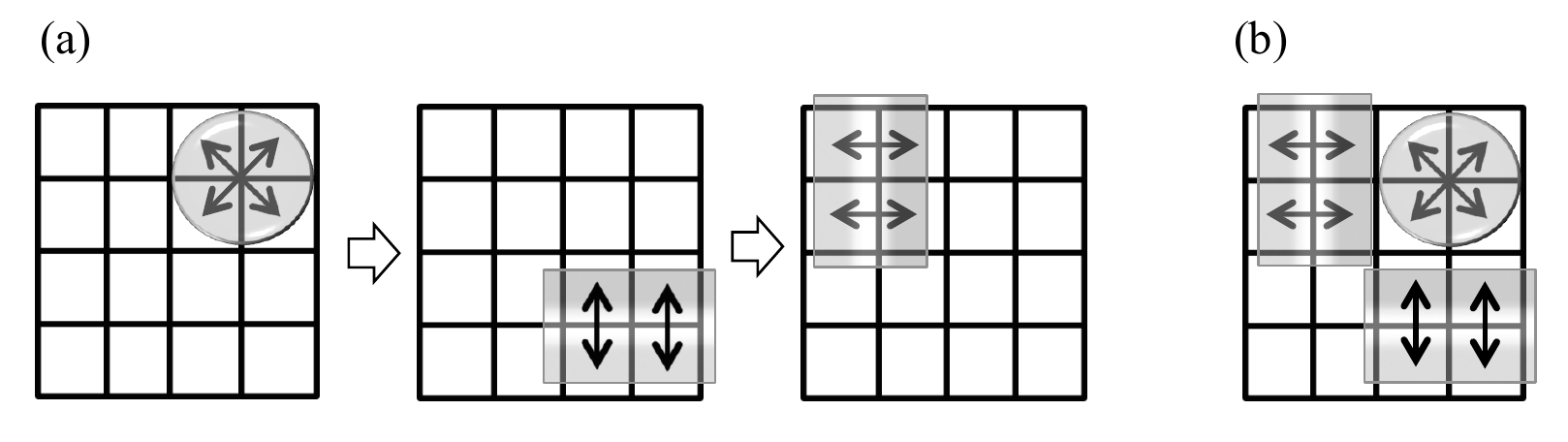}
	\caption{(a) Compact representation of the sequence of pair of confocal lenses representing the CZ gate.(b) Equivalent CZ gate configuration for equal-focus coplanar lenses.}
	\label{FIG22}
\end{figure}

Finally, the implementation of the last gate used in the algorithm, the CNOT gate, is achieved through the decomposition shown in Fig.~\ref{FIG4}. Since we have already discussed the implementation of the Hadamard gate, it remains to show the lenses implementation of the CZ gate. The permutations to realize the CZ gate are shown in Fig.~\ref{FIG22}. These permutations are accomplished by the sequence of pairs of confocal lenses shown in Fig.~\ref{FIG22}(a).

We note that one can use either the same cylindrical lenses employed for the Hadamard gate, taking care to place them in the beam paths of just {\it two} classical bits of the appropriate line and column, or, alternatively, cylindrical lenses with a shorter longitudinal axis, as shown in Fig.~\ref{FIG22}. If the supports of the lenses permit and the foci are equal, the first and the second lenses of the pairs can be mounted in the same planes, as shown in Fig.~\ref{FIG22}(b). This, of course, is not the case of the Pauli-X gate. In the case of the configuration in sequence, similarly as the Pauli-X gate, the spatial order of the pairs is irrelevant.

Since all the necessary gates have been illustrated separately, one can easily obtain the sequence that implements the Deutsch algorithm for the four oracles.

\subsection{Experimental setup}

\begin{figure}[h!]
	\centering
	\includegraphics[scale=0.4]{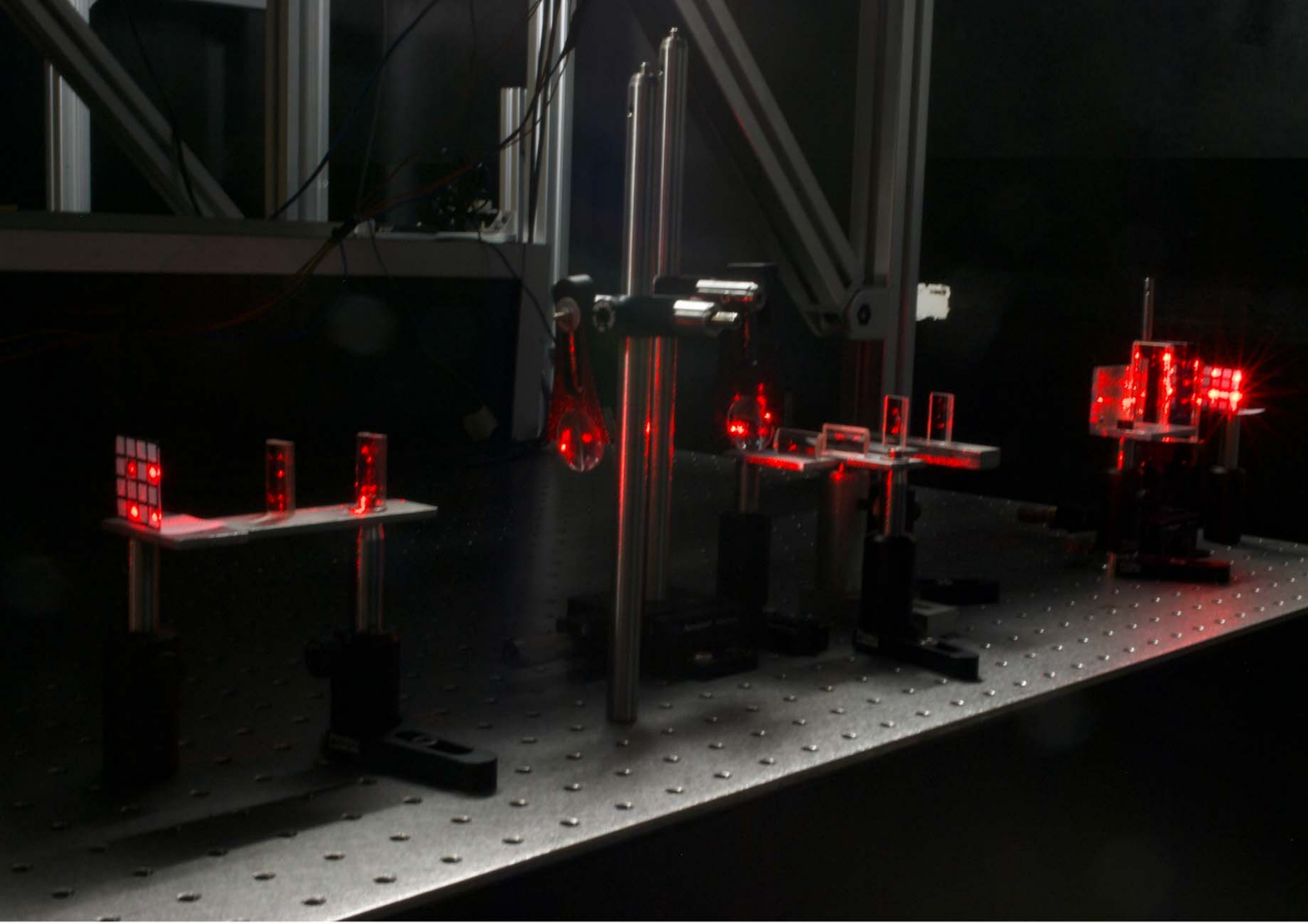}
	\caption{Optical implementation of the Deutsch algorithm for the oracle corresponding to $f(x) = \bar{x}$. The front and back grids represent the initial and final states $|+\rangle|-\rangle$ and $|-\rangle|-\rangle$, respectively.}
	\label{FIG23}
\end{figure}

We will now discuss for illustration, and as a proof of principle, our implementation of the Deutsch algorithm for the oracle $f(x) = \bar{x}$, which is shown in Fig.~\ref{FIG23}. This is one of many possible implementations since the setup admits other configurations as some gates can be put in a different order, provided that they commute. We recommend the setup that best suits the mechanical supports available. The circuit for the remaining oracles are simpler variations of this one.

In our implementation we suppressed the initial Hadamard gates, preparing directly the state $|+\rangle|-\rangle$, as shown in the front grid of Fig.~\ref{FIG23}, since it will be the same no matter which oracle is being interrogated.  The same is true for the final Hadamard, since it is just a matter of interpreting $|+\rangle |-\rangle$ as the result for constant functions and $|-\rangle|-\rangle$ as the result for balanced functions. As the oracle implemented in Fig.~\ref{FIG23} corresponds to a balanced function, the final state shown in the back grid is $|-\rangle|-\rangle$. We also omit lenses whenever we know that no light beams would be incident on them, although in a general case they would be important for the correct realization of the gate. At last, we invert some spatial order of pair of lenses either to minimize aberration and divergence effects or to better adjust the mechanical support system, whenever such inversion would not alter the function being performed. We use 638.2 nm/4.6 mW diode lasers, two types of cylindrical lenses and one type of spherical lenses. We note that there is virtually no restriction on the type of laser to be used and we simply have chosen diode lasers due their low cost and large availability. The  small cylindrical lenses, used to permutate adjacent classical bit lines and columns, have dimensions 12,7mm $\times$ 25,4 mm and focal distance of 19 mm. The large cylindrical lenses, used to permutate non-adjacent classical bits columns have dimensions 25 mm $\times$ 50 mm and focal distance of 25,4 mm. The spherical lenses have diameter of 2,5 mm and focal distance of 50 mm. The side length of the square grid used for this lens configuration is 2.9 cm, which gives, on average, a distance of  0,7 mm between adjacent beam spots. All these dimensions can vary depending on the size of the chosen lenses. 

\begin{figure}[h!]
	\centering
	\includegraphics[scale=0.7]{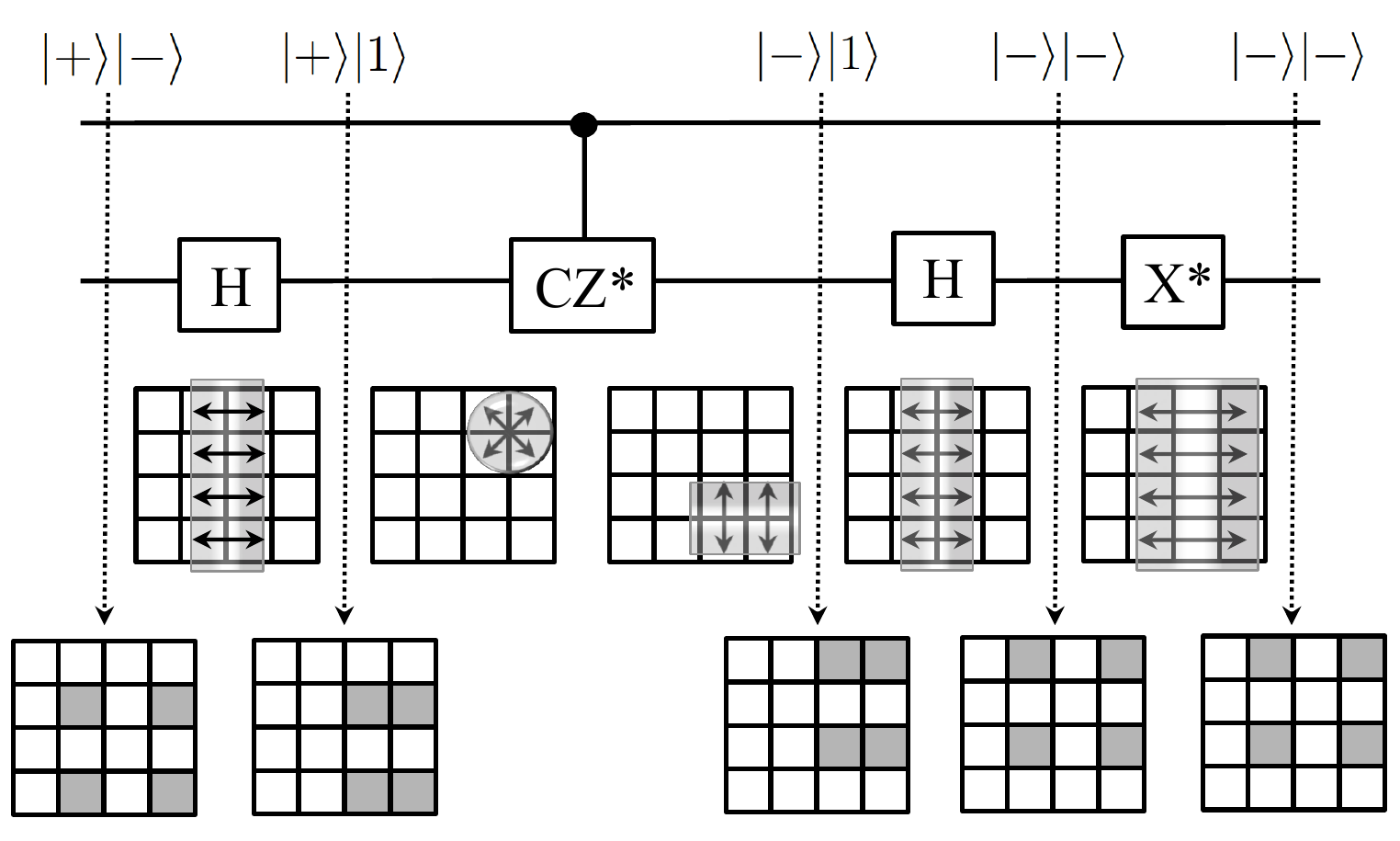}
	\caption{Compact representation of the sequence of lenses of the experimental implementation of the Deutsch algorithm for the oracle corresponding to $f(x) = \bar{x}$.}
	\label{FIG24}
\end{figure}

The schematic sequence of lenses used in this experimental implementation is shown in Fig.~\ref{FIG24} and actually corresponds to a simplified realization of the oracle for the balanced function $f(x) = \bar{x}$. Whenever we omit lenses due the lack of an incident beam we will use an asterisk. 

Note that, since the CNOT gate (represented by the sequence of gates H CZ* H in Fig.~\ref{FIG24}), comutes with the Pauli-X gate, for experimental convenience, we invert the spatial order of the pairs of lenses corresponding to these gates, in comparison with the order displayed in Fig.~\ref{FIG7}(d).

\section{Conclusion}

In this work we introduce some very basic principles of quantum computation and information. The seminal quantum Deutsch algorithm, whose aim is to distinguish between constant and balanced functions, is discussed in details, exemplifying how quantum mechanics properties can be useful for computation. Based on a toy model, we develop an extremely simple optical implementation of a recent proposal for a classical analogue of the quantum Deutsch algorithm using only low-cost devices such as lenses and diode lasers.  Beyond its technical simplicity, this optical implementation has the advantage of allowing  for the visualization of all the intermediate steps of the algorithm.  Inserting a glass grid between pair(s) of lenses, it is possible to follow the evolution of states resulting from the sequential action of the gates. This possibility helps a first understanding on how a quantum circuit works and this comprehension extends naturally to more complex computations. 

We should point out that any time-efficient implementation of a quantum algorithm using classical resources  implies, to the best of our knowledge, in a problem of scalability \cite{Monken, Duzzioni, Konrad}. In our case, one case easily note that the number of classical bits increases exponentially with the number of input qubits, which impacts directly the grids and lenses dimensions. 

Nevertheless, scalability is not relevant for this pedagogical-guided work, whose goal is to bring quantum information and computation accessible for non-specialized readers and therefore contributing to interdisciplinary research. Finally, we hope this article pave the way for the investigation of similar classical implementations of other well-known quantum algorithms, such as the Grover \cite{Grover}  and the Simon \cite{Simon} algorithms, for database search and determination of the period of a function, respectively, as well as quantum information protocols such as teleportation \cite{Bennett} and dense coding \cite{Wiesner}.

\begin{acknowledgments}

We thank Ot\'avio Cals for discussions during the development of this work. The authors acknowledge
financial support from the Brazilian funding agencies
FAPERJ, CNPq, CAPES, and the National Institute of Science
and Technology for Quantum Information.

\end{acknowledgments}

\end{document}